\documentclass[pra,twocolumn,showpacs]{revtex4}
\usepackage{epsfig}
\usepackage{amsbsy}
\usepackage{amsmath}
\usepackage{latexsym}

\begin{document}
\title{Entanglement properties in (1/2,1) mixed-spin Heisenberg systems}
\author{Zhe Sun, XiaoGuang Wang, AnZi Hu, and You-Quan Li}
\affiliation{Zhejiang Institute of Modern Physics, Department of
Physics, Zhejiang University, HangZhou 310027, China}

\date{\today}
\begin{abstract}
By using the concept of negativity, we investigate entanglement in
(1/2,1) mixed-spin Heisenberg systems.  We obtain the analytical
results of entanglement in small isotropic Heisenberg clusters
with only nearest-neighbor (NN) interactions up to four spins and
in the four-spin Heisenberg model with both NN and
next-nearest-neighbor (NNN) interactions. For more spins, we
numerically study effects of temperature, magnetic fields, and NNN
interactions on entanglement. We study in detail the threshold
value of the temperature, after which the negativity vanishes.
\end{abstract}
\pacs{75.10.Jm,03.65.Ud} \maketitle
\section{Introduction}
Entanglement, an essential feature of the quantum mechanics, has
been introduced in many fields of physics. In the field of quantum
information, the entanglement has played a key role. The study of
entanglement properties in many-body systems have attracted much
attention~\cite{M_Nielsen}-\cite{QPT_GVidal}. The Heisenberg
chains, widely studied in the condensed matter field, display rich
entanglement features and have many useful applications such as in
the quantum state transfer~\cite{M_Sub}.

Most of the systems considered in previous studies are spin-half
systems as there exists a good measure of entanglement of two
spin-halves, the concurrence~\cite{Conc}, which is applicable to
an arbitrary state of two spin halves. On the other hand, the
entanglements in mixed-spin or higher spin systems are not
well-studied due to the lack of good operational entanglement
measures. There are several initial studies along this
direction~\cite{Schliemann,Yi,Zhu}, however these works are
restricted to the case of two particles.

For the case of higher spins, a non-entangled state has
necessarily a positive partial transpose according to the
Peres-Horodecki criterion~\cite{PH}. In the case of two spin
halves, and the case of (1/2,1) mixed spins, fortunately, a
positive partial transpose is also sufficient. Thus, the
sufficient and necessary condition for entangled state in (1/2,1)
mixed spin systems is that it has a negative partial transpose.
This allows us to investigate entanglement features of the mixed
spin system.

The Peres-Horodecki criterion give a qualitative way for judging
whether the state is entangled or not. The quantitative version of
the criterion was developed by Vidal and Werner~\cite{Vidal}. They
presented a measure of entanglement called negativity that can be
computed efficiently, and the negativity does not increase under
local manipulations of the system. The negativity of a state
$\rho$ is defined as
\begin{equation}
{\cal N(\rho)}=\sum_i|\mu_i|,
\end{equation}
where $\mu_i$ is the negative eigenvalue of $\rho^{T_1}$, and $T_1$ denotes \\
the partial transpose with respect to the first system. The
negativity ${\cal N}$ is related to the trace norm of $\rho^{T_1}$
via
\begin{equation}
{\cal N(\rho)}=\frac{\|\rho^{T_1}\|_1-1}{2},
\end{equation}
where the trace norm of $\rho^{T_1}$ is equal to the sum of the
absolute values of the eigenvalues of $\rho^{T_1}$. In this paper,
we will use the concept of negativity to study entanglement in
(1/2,1) mixed-spin systems.

As shown in most previous works, models with the NN exchange
interactions are considered and it is not easy to have pairwise
entanglement between the NNN spins~\cite{M_Osborne,M_Osterloh}. It
is true that there exist some quasi-one-dimension compounds
offering us systems with NNN interactions. Bose and
Chattopadhyay~\cite{Ibose} and Gu et al.~\cite{Gu} have
investigated entanglement in spin-half Heisenberg chain with NNN
interactions. In our paper here, we study entanglement properties
not only in the (1/2,1) mixed-spin systems only with NN
interactions, but also in the system with NNN interactions.

Entanglement in a system with a few spins displays general
features of entanglement with more spins. For instance, in the
anisotropic Heisenberg model with a large number of qubits, the
pairwise entanglement shows a maximum at the isotropic
point~\cite{Gu}. This feature was already shown in a small system
with four or five qubits~\cite{Wang04}. So, the study of small
systems is meaningful in the study of entanglement as they may
reflect general features of larger or macroscopic systems. Also,
due to the limitation of our computation capability, we only
concentrate on small systems such as 4, 5 and 6-spin models.

The paper is organized as follows. In Sec. II, we study the
systems with only NN interactions. The analytical results of
negativity for the cases of two and three spins are given. The
relation between entanglement and the macroscopic thermodynamical
function, the internal energy is revealed. Also we numerically
compute the negativity in more general mixed-spin models up to
eight spins, and consider the effects of magnetic fields in this
section. In Sec. III, the system with NNN interaction is
discussed. For the four-spin case, we analytically calculate the
eigenenergy of the system from which we get the analytical results
of the negativity of the NN spins. We numerically study
negativities versus NNN exchanging coupling, and the case of
finite temperature is also considered. For larger system up to
eight-spin system, we get some numerical results. The conclusion
is given in Sec. IV.
\section{Entanglement in Heisenberg chain only with nearest-neighbor interaction}
\subsection{Analytical results of Entanglement in Heisenberg
models} We study entanglement of states of the system at thermal
equilibrium described by the density operator $\rho(T)=\exp(-\beta
H)/Z$, where $\beta=1/k_BT$, $k_B$ is the Boltzmann's constant,
which is assumed to be one throughout the paper , and
$Z=\text{Tr}\{\exp(-\beta H)\}$ is the partition function. The
entanglement in the thermal state is called thermal entanglement.

We consider two kinds of spins, spin $\frac{1}{2}$ and $1$,
alternating on a ring with antiferromagnetic exchange coupling.
The Hamiltonian is given by
\begin{eqnarray}
H_0&=&\sum_{i=1}^{N/2} ({\bf s}_i\cdot{\bf S}_{i}+{\bf S}_{i}\cdot
{\bf
s}_{i+1}), (N\in \text{even})\label{H01}\\
 H_0&=&\sum_{i=1}^{(N-1)/2} ({\bf s}_i\cdot{\bf S}_{i}+{\bf S}_{i}\cdot {\bf s}_{i+1})+
 {\bf s}_{\frac{(N+1)}{2}}\cdot{\bf s}_{1},\nonumber\\
 &&(N\in \text{odd}),\label{H02}
\end{eqnarray}
where ${\bf s}_i$ and ${\bf S}_i$ are spin-1/2 and spin-1
operators, respectively. The exchange interactions exist only
between nearest neighbors, and they are of the same strength which
are set to one. We adopt the periodic boundary condition. In
Fig.~1, we give the schematic representation of the above
Hamiltonian. Next, we first consider the models with two and three
spins, and aim at getting analytical results of entanglement.
\subsubsection{Two-spin case}
For the two-spin case, the Hamiltonian~(\ref{H01}) reduces to
$H_0={\bf s}_1\cdot{\bf S}_1$. To have a matrix representation of
the Hamiltonian, we choose the following basis
\begin{equation}
\Big\{ |-\frac{1}{2},-1\rangle,  |\frac{1}{2}, 0\rangle,
|-\frac{1}{2}, 1\rangle, |\frac{1}{2} , -1\rangle, |-\frac{1}{2},
0\rangle, |\frac{1}{2} ,1\rangle\Big\},
\end{equation}
where $|m,M\rangle$ is the eigenstate of ${s}_z$ and $S_z$ with
the corresponding eigenvalues given by $m$ and $M$, respectively.
\begin{figure}
\includegraphics[width=0.35\textwidth]{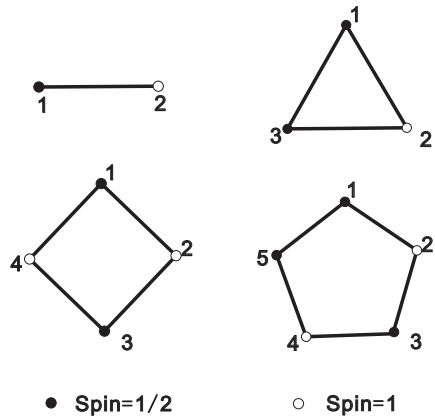}
\caption{Schematic representation of the mixed spin Hamiltonian
for the number of spins from 2 to 5.}
\end{figure}

In the above basis, the Hamiltonian can be written as a
block-diagonal form with the dimension of each block being at most
$2\times 2$. Thus, the density matrix $\rho_{12}$ for the thermal
state is obtained as
\begin{equation}
\rho_{12}=\left(
\begin{array}{llllll}
a_1&0 & 0 &0 & 0 & 0\\
0 &a_2&b_1&0 & 0 & 0 \\
0& b_1&a_3&0 & 0 & 0 \\
0& 0  &0  &a_4&b_2&0\\
0& 0 & 0 & b_2&a_5&0\\
0& 0 & 0 & 0 & 0 & a_6
\end{array}
\right) , \label{rho}
\end{equation}
with the partition function and the matrix elements given by
\begin{eqnarray}
Z&=&2 e^\beta+4e^{-\frac{1}{2}\beta},\label{para1}\\
a_1&=&a_6=e^{-\frac{1}{2}\beta}/Z,\label{para22}\\
a_2&=&a_5=\frac{1}{6},\label{para222}\\
a_3&=&a_4=\frac{1}3\left(2e^\beta+e^{-\frac{1}{2}\beta}\right)/Z,\\
b_1&=&b_2=\frac{\sqrt{2}}{3}\left(e^{-\frac{1}{2}\beta}-e^\beta\right)/Z,\nonumber\\
&=&\sqrt{2}(a_1-a_2). \label{para2}
\end{eqnarray}
After the partial transpose with respect to the first spin-half
subsystem, we can get $\rho_{12}^{T_1}$
\begin{equation}
\rho_{12}^{T_1}=\left(
\begin{array}{llllll}
a_1&b_2&0  &0  &0  &0\\
b_2&a_2&0  &0  &0  &0\\
0  &0  &a_3&0  &0  &0\\
0  &0  &0  &a_4&0  &0\\
0  &0  &0  &0  &a_5&b_1\\
0  &0  &0  &0  &b_1&a_6
\end{array}
\right),  \label{rhoT1}
\end{equation}
which is still of the block-diagonal form, and computation of its
eigenvalues is straightforward. There are only two eigenvalues
which are possibly negative. The negativity is thus given by
\begin{align}
{\cal N}(\rho_{12})=&\frac{1}2\max \big[0,\sqrt{(a_1-a_2)^2+4b_2^2}-a_1-a_2)\big]\nonumber\\
+&\frac{1}2\max\big[0,\sqrt{(a_5-a_6)^2+4b_1^2}-a_5-a_6\big].\label{N}
\end{align}

Substituting Eqs.~(\ref{para1})--(\ref{para2}) leads to the
analytical result of negativity
\begin{align}
{\cal N}(\rho)=&\max
\big[0,\sqrt{(a_1-a_2)^2+4b_2^2}-a_1-a_2)\big]\nonumber\\
=&2\max[0,a_2-2a_1]\nonumber\\
=&\frac{1}{3}\max\Big[0,\frac{e^\beta-4e^{-\frac{1}{2}\beta}}
{e^\beta+2e^{-\frac{1}{2}\beta}}\Big],\label{N1}
\end{align}
where the second equality follows from Eq.~(\ref{para2}).

We can see that the negativity is a function of the single
parameter $\beta$. In the limit of $T\rightarrow 0$, the
negativity becomes 1/3 and the ground state is entangled. From
Eq.~(\ref{N1}), it is direct to check that the negativity is a
monotonically decreasing function when temperature increases.
After a certain threshold value of the temperature, the
entanglement disappears. This threshold value $T_\text{th}$ can be
obtained as
\begin{equation}
T_\text{th}=3/(4\ln2)\approx 1.0820.
\end{equation}
For a ring of spin-half particles interacting via the Hisenberg
Hamiltonian, it was shown that the pairwise thermal entanglement
is determined by the internal energy~\cite{WangPaolo}. It is
natural to ask if similar relations exist in the present
mixed-spin system. The internal energy can be obtained from the
partition function as
\begin{equation}
U=-\frac{1}{Z}\frac{\partial Z}{\partial\beta}.
\end{equation}
Substituting Eq.~(\ref{para1}) into the above equation leads to
\begin{equation}
U=\frac{\displaystyle
-e^\beta+e^{-\frac{1}{2}\beta}}{\displaystyle
e^\beta+2e^{-\frac{1}{2}\beta}}. \label{U}
\end{equation}
From Eqs.~(\ref{N1}) and (\ref{U}), we obtain a quantitative
relation between the negativity and the internal energy
\begin{equation}
{\cal N(\rho)}=\frac{1}{3}\max[0,-1-2U].\label{U1}
\end{equation}
The above equation builds a connection between the microscopic
entanglement and the macroscopic thermodynamical function, the
internal energy. The internal energy completely determine the
thermal entanglement. From the equation, we can also read that the
thermal state becomes entangled if and only if the internal energy
$U<-1/2$. Since $U=\langle H\rangle=\langle \bf s_1\cdot\bf
S_1\rangle$, we have
\begin{equation}
{\cal N(\rho)}=\frac{1}{3}\max[0,-1-2\langle\bf s_1\cdot\bf
S_1\rangle],\label{U2}
\end{equation}
which is consistent with the result obtained in
Ref.~\cite{Schliemann} by the group-theoretical technique.
\subsubsection{Three-spin case}
We now consider the three-spin case and the schematic
representation of the corresponding Hamiltonian is given by
Fig.~1. In this situation, there are two types of pairwise
entanglement, the entanglement between spin 1/2 and spin 1 and the
entanglement between two spin halves.

The eigenvalue problem can be solved analytically, and after
tracing out the third spin-half system the reduced density matrix
$\rho_{12}$ is still of the same form as in Eq.~(\ref{rho}) with
matrix elements given by
\begin{eqnarray}
a_1&=&\left(\frac{5}{4}e^{-\frac{5}{4}\beta}+\frac{3}{4}e^{\frac{3}{4}\beta}\right)/Z,\nonumber\\
a_2&=&\left(\frac{5}{6}e^{-\frac{5}{4}\beta}+e^{\frac{3}{4}\beta}+\frac{1}{6}e^{\frac{7}{4}\beta}\right)/Z,\nonumber\\
a_3&=&\left(\frac{5}{12}e^{-\frac{5}{4}\beta}+\frac{5}{4}e^{\frac{3}{4}\beta}+\frac{1}{3}e^{\frac{7}{4}\beta}\right)/Z,\nonumber\\
b_1&=&\left(\frac{5\sqrt{2}}{12}e^{-\frac{5}{4}\beta}-\frac{\sqrt{2}}{4}e^{\frac{3}{4}\beta}-\frac{\sqrt{2}}{6}e^{\frac{7}{4}\beta}\right)/Z,\nonumber\\
Z&=&5e^{-\frac{5}{4}\beta}+6e^{\frac{3}{4}\beta}+e^{\frac{7}{4}\beta}.
\end{eqnarray}
Substituting the above equations to Eq.~(\ref{N}) leads to the
negativity
\begin{equation}
{\cal N}(\rho_{12})=\max\Big[0,
{-\frac{10}{3}e^{-\frac{5}{4}\beta}-e^{\frac{3}{4}\beta}+\frac{1}{3}e^{\frac{7}{4}\beta}}\Big]/Z.
\end{equation}
It is evident that the negativity becomes 1/3 in the limit of
$T\rightarrow 0$. From the expression of the negativity, the
threshold value of temperature after which the entanglement
vanishes can be estimated as
\begin{equation}
T_{\text{th}}\approx 1/\ln3.2719\approx0.7609.
\end{equation}

To examine the entanglement between two spin halves, we trace out
the spin 1 system and get the reduced density matrix $\rho_{13}$
as follows
\begin{equation}
\rho_{13}=\left(
\begin{array}{llll}
a_1& 0&  0&   0\\
0&   a_2&b&   0\\
0&   b&  a_2& 0\\
0&   0&  0&   a_1
\end{array} \right),\label{rho13}
\end{equation}
with the matrix elements given by
\begin{eqnarray}
a_1&=&\left(\frac{5}{3}e^{-\frac{5}{4}\beta}+e^{\frac{3}{4}\beta}+\frac{1}{3}e^{\frac{7}{4}\beta}\right)/Z\nonumber\\
a_2&=&\left(\frac{5}{6}e^{-\frac{5}{4}\beta}+2e^{\frac{3}{4}\beta}+\frac{1}{6}e^{\frac{7}{4}\beta}\right)/Z\nonumber\\
b&=&\left(\frac{5}{6}e^{-\frac{5}{4}\beta}-e^{\frac{3}{4}\beta}+\frac{1}{6}e^{\frac{7}{4}\beta}\right)/Z.
\label{ele}
\end{eqnarray}
After taking the partial transpose, we can get $\rho^T_{13}$
\begin{equation}
\rho^T_{13}=\left(
\begin{array}{llll}
a_1& 0&  0&   b\\
0&   a_2&0&   0\\
0&   0&  a_2& 0\\
b&   0&  0&   a_1
\end{array} \right).\label{rhoT13}
\end{equation}
Then, the negativity is readily obtained as
\begin{equation}
{\cal N}(\rho_{13})=\max[0,|b|-a_1].
\end{equation}
It is straightforward to check that the negativity is always zero.
Or, from another way, all the eigenvalues of the matrix
$\rho^T_{13}$ are obtained as
\begin{eqnarray}
\lambda_1&=&\left(\frac{5}{2}e^{-\frac{5}{4}\beta}+\frac{1}{2}e^{\frac{7}{4}\beta}\right)/Z,\nonumber\\
\lambda_2&=&\left(\frac{5}{6}e^{-\frac{5}{4}\beta}+2e^{\frac{3}{4}\beta}+\frac{1}{6}e^{\frac{7}{4}\beta}\right)/Z,\nonumber\\
\lambda_3&=&\lambda_4=a_2.
\end{eqnarray}
Obviously the negativity vanishes here, in other words there is no
entanglement between the two spin halves.

The ground-state negativity ${\cal N}_{12}={\cal
N}(\rho_{12})$=1/3 and ${\cal N}_{13}={\cal N}(\rho_{13})$=0, here
${\cal N}_{12}$ denotes the negativity between the $1_\text{th}$
and $2_\text{th}$ spin on the chain and ${\cal N}_{13}$ denotes
the one between the $1_\text{th}$ and $3_\text{th}$ spin. The
equation above can be obtained from the non-degenerate ground
state given by:
\begin{eqnarray}
|\psi_0\rangle&=&\frac{\sqrt{6}}{6}\left(|\frac{1}{2},0,-\frac{1}{2}\rangle+|-\frac{1}{2},0,\frac{1}{2}\rangle\right)\nonumber\\
&-&\frac{\sqrt{3}}{3}\left(|\frac{1}{2},-1,\frac{1}{2}\rangle+|-\frac{1}{2},1,-\frac{1}{2}\rangle\right).
\end{eqnarray}
It is interesting to see that the ground-state entanglement
between the spin half and spin 1 in the three-spin case is the
same as that in the two-spin case.

Due to the SU(2) symmetry in our system, there are following
relations between correlation functions and
negativities~\cite{Schliemann}
\begin{align}
{\cal N}_{12}=&-\frac{1}3-\frac{2}3\langle{\bf s}_1\cdot{\bf
S}_1\rangle,\label{N12}\nonumber\\
{\cal N}_{23}=&-\frac{1}3-\frac{2}3\langle{\bf S}_1\cdot{\bf
s}_2\rangle,\nonumber\\
{\cal N}_{31}=&-\frac{1}4-\langle{\bf s}_2\cdot{\bf s}_1\rangle,
\end{align}
where we have removed the max function in the negativity, implying
that the negative value of ${\cal N}$ indicates no entanglement.
Then, we have the relation between the internal energy and the
negativities
\begin{align}
U=&-\frac{5}4-{\cal N}_{13}-\frac{3}2({\cal N}_{12}+{\cal N}_{23})
\nonumber\\
=&-\frac{5}4-{\cal N}_{13}-{3}{\cal N}_{12}. \label{relation}
\end{align}
The second equality follows from the exchange symmetry, namely,
the Hamiltonian is invariant when exchanging two spin halves. So,
for the three-spin case, the internal energy is related to two
negativities.

To apply the above result, we consider the the ground-state
properties ($T=0$). The Hamiltonian can be rewritten as
\begin{equation}
H=\frac{1}{2}[({\bf s}_1+{\bf S}_1+{\bf s}_2)^2-{\bf s}_1^2-{\bf
S}_1^2-{\bf s}_2^2].
\end{equation}
Then, by the angular momentum coupling theory, the ground-state
energy is obtained as $E_0=-7/4$. Substituting the ground-state
energy and ${\cal N}_{12}=1/3$ to Eq.~(\ref{relation}), we obtain
${\cal N}_{13}=-1/2$, indicating that there exists no entanglement
between two spin halves. Next, we consider more general
situations, i.e., the case of even $N$ sites.

\subsubsection{The case of even $N$ spins} Except for the SU(2)
symmetry in the system, there exists exchange symmetry for the
case of even spins. For instance, in the four-spin model, the
Hamiltonian is invariant when exchanging two spin halves or two
spin ones. Thus, for even-spin model, the entanglements between
the two nearest-neighbor spins and the correlation functions
$\langle{\bf s}_i\cdot{\bf S}_{i}\rangle$ are independent on index
$i$. Therefore, the internal energy per spin is equal to the
correlation function $\langle{\bf s}_i\cdot{\bf S}_{i}\rangle$
\begin{equation}
u=U/N=\langle{\bf s}_1\cdot{\bf S}_{1}\rangle. \label{uuu}
\end{equation}
From Eqs.~(\ref{N12}) and (\ref{uuu}), we have
\begin{equation}
{\cal N}_{12}=-\frac{1}3-\frac{2}3 u. \label{N1212}
\end{equation}
This equation indicates that for the case of even spins the
entanglement between two nearest neighbors is solely determined by
the internal energy per spin. And for the case of zero
temperature, the entanglement is determined by the ground-state
energy. The less the energy, the more the entanglement.

We now apply the above result to the study of ground-state
entanglement. We rewrite the four-spin Hamiltonian as follows
\begin{equation}
H=\frac{1}{2}[({\bf s}_1+{\bf S}_1+{\bf s}_2+{\bf S}_2)^2-({\bf
s}_1+{\bf s}_2)^2-({\bf S}_1+{\bf S}_2)^2].
\end{equation}
Then, by the angular momentum coupling theory, we obtain the
ground-state energy per site $e_0=E_0/4=-3/4$. Thus, from
Eq.~(\ref{N1212}), we have ${\cal N}_{12}=1/6$. For $N\ge 5$, it
is hard to get analytical results. We next numerically calculate
the entanglement for the case of more spins, and also consider the
effects of magnetic fields.

\subsection{Numerical results} Having
obtained analytical results of entanglement in the Heisenberg
model with a few spins, we now numerically examine the
entanglement behaviors in more general Hamiltonian including more
spins and magnetic fields.

\subsubsection{Entanglement versus temperature}
We consider the entanglement versus temperature for different
number of spins $N$, and the numerical results are plotted in
Fig.~2 and Fig.~3. It is clear to see that the ground state
exhibits maximal entanglement, and with the increase of
temperature, the entanglement monotonically decreases until it
reaches zero. The decrease of entanglement is due to the mixture
of less entangled excited states when increasing the temperature.
The existence of the threshold temperature is also evident. For
the case of even number of sites (Fig.~2), the threshold
temperature decreases with the increase of $N$. In contrast to
this behavior, for the case of odd number of spins, as seen from
Fig.~3, the threshold temperature increases with the increase of
$N$.
\begin{figure}
\includegraphics[width=0.45\textwidth]{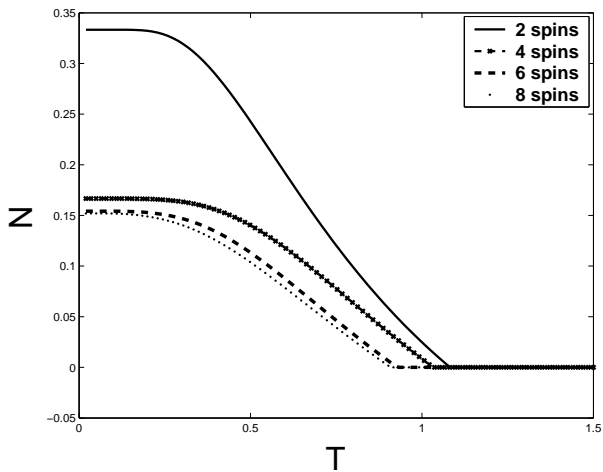}
\caption{ Negativity versus temperature for different even number
of spins in the Heisenberg model.}
\end{figure}

\begin{figure}
\includegraphics[width=0.45\textwidth]{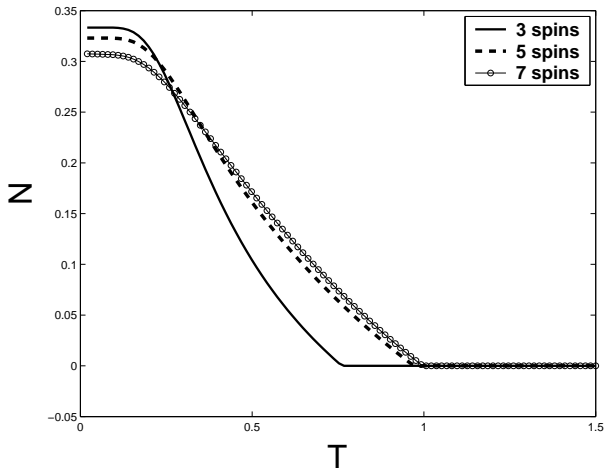}
\caption{ Negativity versus temperature for different odd number
of spins in the Heisenberg model.} \
\end{figure}

\subsubsection{Effects of magnetic fields}
We now examine the effect of magnetic fields on entanglement. The
Heisenberg Hamiltonian with a magnetic field along $z$ direction
is given by
\begin{equation}
H_1=H_0+B\sum_{i=1}^{N/2} \big({s}_{iz}+{S}_{i,z}\big), (N\in
\text{even}).
\end{equation}
For $N=2$, the analytical result of negativity can be obtained
(see Appendix A).

\begin{figure}
\includegraphics[width=0.45\textwidth]{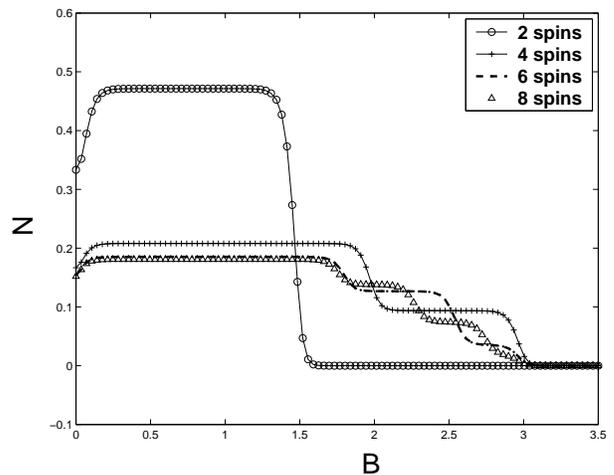}
\caption{Negativity versus the magnetic field for $T=0.05$.} \
\end{figure}

In Fig.~4, we plot the negativity versus the magnetic field at a
low temperature. For the two-spin case, with the increase of the
magnetic field, the negativity rapidly reach a platform, and then
after a certain magnetic field $B_\text{th}$, it jumps down to
zero, indicating no entanglement. When $B=0$, the ground state is
two-fold degenerate with ${\cal N}=1/3$. For a small $B>0$, the
system is no longer degenerate, and the ground state is given by
\begin{equation}
|\psi_0\rangle=\frac{\sqrt{6}}{3}|\frac{1}{2},-1\rangle-\frac{\sqrt{3}}{3}|-\frac{1}{2},0\rangle,
\end{equation}
which is of the Schmidt form. For a state written as its Schmidt
form
\begin{equation}
|\psi\rangle=\sum_n c_n|e_n\rangle|f_n\rangle,
\end{equation}
the negativity is obtained as~\cite{Vidal}
\begin{equation}
{\cal N}=\left[\big(\sum_n c_n\big)^2-1\right]/2.
\end{equation}
Here, $C_n$ are the Schmidt coefficients, and $\{|e_n\rangle\}$
and $\{|f_n\rangle\}$ are bases for subsystems 1 and 2,
respectively. Applying this formula to the ground state, we
immediately have ${\cal N}=\sqrt{2}/3$. When $B=B_\text{th}=3/2$,
the ground state is two-fold degenerate, and when $B>B_\text{th}$
the ground state becomes non-degenerate and the corresponding wave
function is given by
\begin{equation}
|\psi_0\rangle=|-\frac{1}2,-1\rangle,
\end{equation}
which is obviously of no entanglement. Then, the ground-state
negativity forms a platform when $0<B<B_\text{th}$. The jump of
negativity at $B=B_\text{th}$ is due to the level crossing. For
$N>2$, the effects of magnetic fields on entanglement can be also
explained by level crossing. For instance, for $N=4$, there are
two level crossing, and the entanglement displays two jumps.

\section{Effects of next-nearest-neighbor interactions on entanglement}
We have studied the effects of finite temperature and magnetic
fields on entanglement, and now consider the model containing two
kinds of spins, spin $\frac{1}{2}$ and $1$, alternating on a ring
with antiferromagnetic exchange coupling between both the NN spins
and the NNN spins. The Hamiltonian can be expressed as
\begin{equation}
H=J_1\sum^{N/2}_{i=1}\big({\bf s}_i\cdot {\bf  S}_{i}+ {\bf
S}_{i}\cdot {\bf s}_{i+1}\big)+J_2\sum^{N/2}_{i=1}\big({\bf
s}_i\cdot {\bf  s}_{i+1}+ {\bf S}_{i}\cdot {\bf S}_{i+1}\big),
\end{equation}
where the ${\bf s}_i$ and ${\bf S}_i$ are spin-1/2 and spin-1
operators in the $i$th cell. $J_1$ characterizes the NN exchange
coupling and $J_2$  the NNN coupling. We consider the
antiferromagnetic interaction by taking $J_1,J_2>0$. $N$ is the
total number of spins and here we choose it be even. Also we adopt
the periodic boundary condition.

\subsection{Four-spin model}
\subsubsection{Eigenenergy and ground-state entanglement} The model
with four spins is the simplest model with  NNN interactions. We
first solve the eigenvalue problem of this model. The key step is
to write the four-spin Hamiltonian in the following form,
\begin{eqnarray}\label{H4}
H&=&\frac{1}{2}\big\{J_1{\bf S}^2+(2J_2-J_1)[({\bf s}_1+{\bf
s}_2)^2+({\bf S}_1+{\bf S}_2)^2]\nonumber\\
&-&2J_2({\bf s}_1^2+{\bf s}_2^2+{\bf S}_1^2+{\bf S}_2^2 )\big\},
\end{eqnarray}
where  ${\bf S}={\bf S}_1+{\bf S}_2+{\bf s}_1+{\bf s}_2$ denotes
the total spin. From the above form, by angular momentum coupling
theory, one can readily obtain all eigenvalues of the system as
\begin{eqnarray}
E_1&=&J_2n(n+1)-\frac{11}{2}J_2,\nonumber\\
E_2&=&nJ_1+n(n+1)J_2-\frac{7}{2}J_2,\nonumber\\
E_3&=&-J_1+n(n+1)J_2-\frac{7}{2}J_2,\nonumber\\
E_4&=&-(n+1)J_1+n(n+1)J_2-\frac{7}{2}J_2, \label{E}
\end{eqnarray}
where parameter $n=0,1,2$ in expressions $E_1$ and $E_2$ and
$n=1,2$ in expressions $E_3$ and $E_4$, respectively. Then, from
Eq.~(\ref{E}), we may find the ground-state energy as
\begin{equation}\label{Egs}
E_\text{GS}=\left\{
\begin{array}{ll}
-3J_1+\frac{5}{2}J_2 &\; \text{when}\;\;\; 0\leq J_2<J_1/4, \\
-2J_1-\frac{3}{2}J_2 &\; \text{when}\;\;\; J_1/4<J_2< J_1/2, \\
-\frac{11}{2}J_2     &\;\text{when} \;\;\;J_2>J_1/2,
\end{array}
\right.
\end{equation}
Clearly, there are two level-crossing points, which will greatly
affect behaviors of ground-state entanglement.

From Eq.~(\ref{H4}), it is obvious that in addition to the the
SU$(2)$ symmetry, there also exists an exchange symmetry, namely,
exchanging two spin halves or two spin ones yields invariant
Hamiltonian. Then, the correlator  $\langle {\bf s}_i\cdot{\bf
S}_i \rangle$ between any NN spins are the same. Therefore,  we
can get the correlator $\langle {\bf s}_i\cdot{\bf S}_i \rangle$
from the ground-state energy via the Hellmann-Feynman theorem.

When $0\leq J_2<J_1/4$, after applying the Hellmann-Feynman
theorem to the ground-state energy, we obtain
\begin{equation}
\langle{\bf s}_1\cdot{\bf S}_1\rangle=\frac{1}{4}\frac{\partial
E_0} {\partial J_1}=-\frac{3}{4}.
\end{equation}
Due to the SU$(2)$ symmetry in our system, we have the following
relation between the negativity and the correlator $\langle{\bf
s}_1\cdot{\bf S}_1\rangle$\cite{Schliemann},
\begin{equation}
{\cal N}_{1/2,1}=\max\big\{0, -\frac{1}3-\frac{2}3\langle{\bf
s}_1\cdot{\bf S}_1\rangle \big \},\label{cor}
\end{equation}
where we use ${\cal N}_{1/2,1}$ to denote the negativity between
the NN spins.  Thus we obtain
\begin{equation}
{\cal N}_{1/2,1}=1/6.
\end{equation}
In other regions, we find that ${\cal N}_{1/2,1}=0$ for
$J_2>J_1/4$.

From the above analytical results, we can find that the negativity
between the NN spins is not a continuous function of parameter
$J_2$. It jumps from the value $\frac{1}{6}$ down to zero at
$J_2=0.25$. Hence, we can see that the NNN interaction may
deteriorate the entanglement of the NN spins. The critical point
of $J_2=0.25$ may be considered to be a threshold value, after
which the NN entanglement vanishes.

Having studied ground-state entanglement, now we make a short
discussion of entanglement of excited states. We consider the
first excited state. In the region $J_1/4<J_2<3J_1/8$, the first
excited energy is given by
\begin{equation}
E_1=-3J_1+\frac{5}{2}J_2,
\end{equation}
from that we can get $\langle{\bf s}_1\cdot{\bf S}_1\rangle=-3/4$
and $N_{1/2,1}=1/6$. While in the rest region the negativity of
the first excited state is zero.
\subsubsection{Thermal entanglement}

Next, we consider the thermal entanglement. {}From Eq.~(\ref{E}),
the partition function can be obtained as
\begin{eqnarray}\label{PF}
Z&=&5e^{-\frac{1}{2}\beta J_2}+6e^{\frac{7}{2}\beta
J_2}+e^{\frac{11}{2}\beta J_2}\nonumber\\
&+&e^{-\frac{5}{2}\beta J_2}(7e^{-2\beta J_1}+5e^{\beta J_1}+3e^{3\beta J_1})\nonumber\\
 &+&e^{\frac{3}{2}\beta J_2}(5e^{-\beta J_1}+3e^{\beta J_1}+e^{2\beta
 J_1}),
\end{eqnarray}
The correlator $\langle{\bf s_1}\cdot {\bf S_1} \rangle$ at finite
temperature can be computed from the partition function via the
following relation
\begin{equation}\label{r1}
\langle{\bf s_1}\cdot {\bf S_1} \rangle=-\frac{1}{4\beta
Z}\frac{\partial Z}{\partial J_1},
\end{equation}

Substituting (\ref{PF}) to Eq.~(\ref{r1}) yields
\begin{eqnarray}
\langle{\bf s_1}\cdot {\bf S_1} \rangle&=&-\frac{1}{4Z}\big [
e^{-\frac{5}{2}\beta J_2}(-14e^{-2\beta J_1 }+5e^{\beta J_1}+9e^{3\beta J_1})\nonumber\\
&+&e^{\frac{3}{2}\beta J_2}(-5e^{-\beta J_1}+3e^{\beta
 J_1}+2e^{2\beta J_1})\big ].
\end{eqnarray}
After substituting the above equation into (\ref{cor}), we may get
analytical expression of the negativity ${\cal N}_{1/2,1}$ at
finite temperatures. The negativity is a function of $J_1$, $J_2$
and $T$.

\textit{Low-temperature case}:
\begin{figure}
\includegraphics[width=0.45\textwidth]{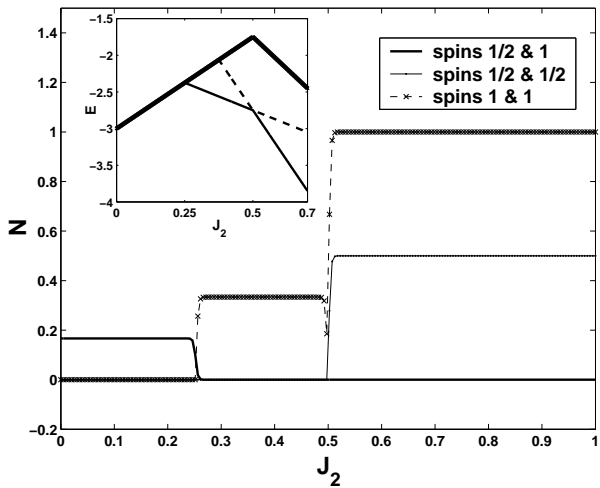}
\caption{Negativity  versus the exchange interaction $J_2$ at the
temperature $T=0.008$ in the four-spin system. The ground-state
and the first two excited-state energy levels versus $J_2$ are
inserted.}
\end{figure}
We now make numerical calculations of entanglement and first
consider the low-temperature case. We take $J_1=1$ in all the
following plots. In our system there exist three kinds of
negativity, the negativity ${\cal N}_{1/2,1}$ between spin-1 and
spin-1/2, ${\cal N}_{1/2,1/2}$ between two spin-1/2, and ${\cal
N}_{1,1}$ between two spin-1.

In Fig.~5, we plot the negativity versus $J_2$ in four-spin system
at a low temperature of $T=0.008$. It is clear to see that ${\cal
N}_{1/2,1}$ keeps a value about $1/6$ when $J_2$ increases from
zero until it reaches the critical point, at which the ${\cal
N}_{1/2,1}$ displays a jump to zero. This behavior of entanglement
is consistent with that at zero temperature from the analytical
results. It is natural to see that increase of NNN exchange
interaction will suppress the entanglement of NN spins, and at
last completely erase the entanglement.

In comparison with ${\cal N}_{1/2,1}$, the negativities ${\cal
N}_{1,1}$ and ${\cal N}_{1/2,1/2}$ behave distinctly. We see that
near the point of $J_2=1/4$, ${\cal N}_{1,1}$ increases quickly to
a steady value about $1/3$, and when $J_2$ reaches the value about
$1/2$, ${\cal N}_{1,1}$ jumps another steady value 1. These two
jumps result from the two level crossing as seen clearly from the
figure inserted. The second level crossing also leads to a small
dip in the curve of ${\cal N}_{1,1}$. The negativity ${\cal
N}_{1/2,1/2}$ displays a jump to a steady value of 1/2 near
$J_2=1/2$. It is evident that the entanglement between NNN spins
is enhanced by increasing NNN interactions. The competition
between NN and NNN interactions leads to rich behaviors of quantum
entanglement. Another observation is that there is a range of
$J_2$, at which negativities ${\cal N}_{1/2,1}$ and ${\cal
N}_{1/2,1/2}$ are zero, and only ${\cal N}_{1,1}$ is not zero.
This indicates that the NNN interaction must be strong enough to
build up the entanglement of two spin halves.

\textit{Entanglement versus $J_2$ and $T$}: As temperature
increases the entanglement will decrease due to the mixing of less
entangled excited states to the thermal state. It is obvious that
there exists a threshold temperature after which the negativity is
zero. In the frustrated system, there exists the parameter  $J_2$,
and with its increase, the negativity ${\cal N}_{1/2,1}$ will
decrease to zero, while ${\cal N}_{1,1}$ and ${\cal N}_{1/2,1/2}$
increase from zero to their maxima. So it is clear that there also
exists a threshold $J_\text{2th}$ corresponding to the boundary
between zero and nonzero negativities.

\begin{figure}
\includegraphics[width=0.45\textwidth]{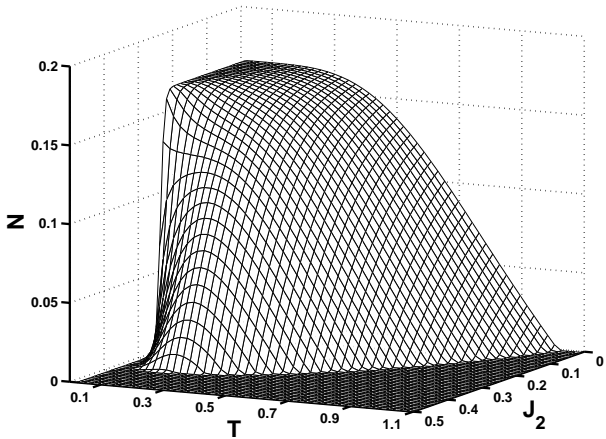}
\caption{Negativity ${\cal N}_{1/2,1}$ versus both the exchange
interaction coefficient $J_2$ and the temperature in the four-spin
system.}
\end{figure}

In  Fig.~6, we show the negativity ${\cal N}_{1/2,1}$ versus the
temperature and $J_2$. When the temperature approaches zero, the
${\cal N}_{1/2,1}$ reaches its maximum, and then with the
temperature increasing, ${\cal N}_{1/2,1}$ decreases to zero. On
the $J_2-T$ plane,  there is a curve along which the negativity
just turns to be zero. It is possible to consider that the curve
describes the threshold $J_\text {2th}$ versus the temperature.
Obviously, $J_\text {2th}$ does not behave as a monotonous
function of the temperature, and it displays a peak at about
$T=0.178$, This behavior is in big contrast with the case of
non-mixed qubit systems~\cite{Gu}. When the temperature rises, the
weight of excited states will increase and it may strongly affect
the negativity. This behavior of $J_\text {2th}$ results from both
the mixture of excited states to the thermal state and the
intrinsic properties of the mixed-spin system. In addition, after
crossing the temperature about $T=1.082$, ${\cal N}_{1/2,1}$ will
vanish, irrespective of the value of $J_2$.

From another point of view, we can read the threshold temperature
$T_\text {th}$ versus different $J_2$ from the curve in the
$J_2-T$ plane. When $J_2$ increases, $T_\text {th}$ decreases, and
when $J_2$ crosses about $0.3758$, ${\cal N}_{1/2,1}$ will
disappear at any temperature.

Next, we consider the entanglement between NNN spins. In Fig.~7,
we plot the negativity  ${\cal N}_{1/2,1/2}$ as a function of the
temperature and $J_2$. We can see that, before $J_2$ reaches the
value about $J_2=0.5$,  ${\cal N}_{1/2,1/2}$ keeps being zero at
any temperature. And in  the region $J_2>0.5$, the ${\cal
N}_{1/2,1/2}$ can be enhanced by the increasing NNN interaction.
This is a result from the competition of two kinds of exchange
interactions. The thermal fluctuation all along suppresses the
entanglement. So, from the curve lying on the $J_2-T$ plane which
corresponds to the boundary of the nonzero and zero values of
${\cal N}_{1/2,1/2}$, we may find that the higher the temperature
is, the larger the threshold $J_\text {2th}$ will be. From another
point of view, the $T_\text {th}$ increases as  $J_2$ increases.
\begin{figure}
\includegraphics[width=0.45\textwidth]{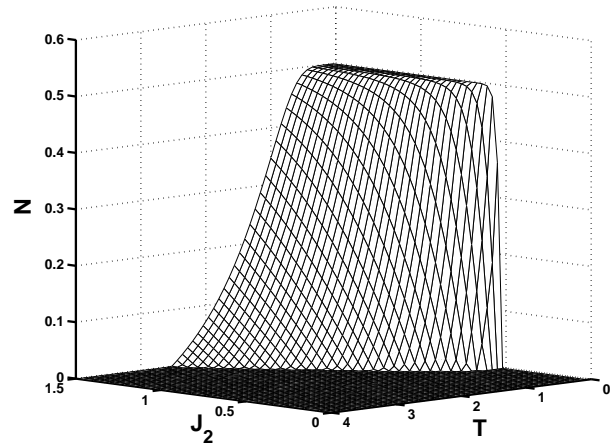}
\caption{Negativity ${\cal N}_{1/2,1/2}$ versus both the exchange
interaction coefficient $J_2$ and the temperature in four-spin
system.}
\end{figure}

In Fig.~8, we plot the negativity ${\cal N}_{1,1}$ versus $T$ and
$J_2$. In the region of $J_2<0.25$, ${\cal N}_{1,1}$ is zero at
any temperature.  When $J_2>0.25$, the increasing NNN exchange
interaction $J_2$ enhances the negativity and exhibits two
particular flat roofs. With the temperature rises,  ${\cal
N}_{1,1}$ is suppressed to zero. Also we can consider the
threshold  $T_\text {th}$ and $J_\text {2th}$ from the critical
curve on the $J_2-T$ plane, and the $J_\text {2th}$ also behaves
as an increasing function of the temperature. We can see the
nonzero region of ${\cal N}_{1,1}$ is much larger than ${\cal
N}_{1/2,1/2}$.   But here it should be pointed out that, because
${\cal N}_{1,1}>0$ only gives a sufficient condition for entangled
state, we can not definitely say that the state in the area of
zero negativity is not entangled.

\begin{figure}
\includegraphics[width=0.45\textwidth]{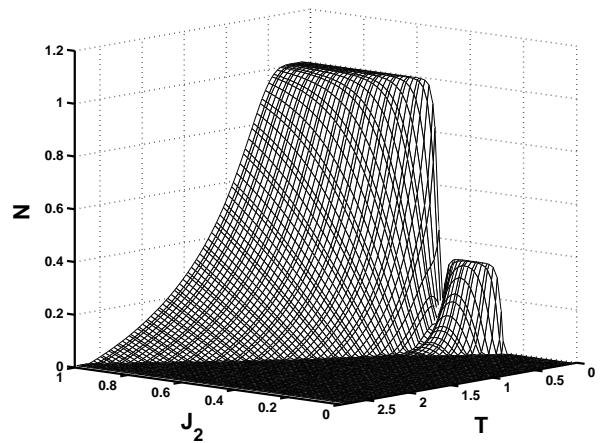}
\caption{Negativity ${\cal N}_{1,1}$ versus both the exchange
interaction coefficient $J_2$ and the temperature in four-spin
system.}\
\end{figure}
\subsection{Numerical results of negativity for more spins} In this
section, we present numerical results of negativity for more
spins, and first consider the low-temperature case.
\subsubsection{Low-temperature case}
\begin{figure}
\includegraphics[width=0.45\textwidth]{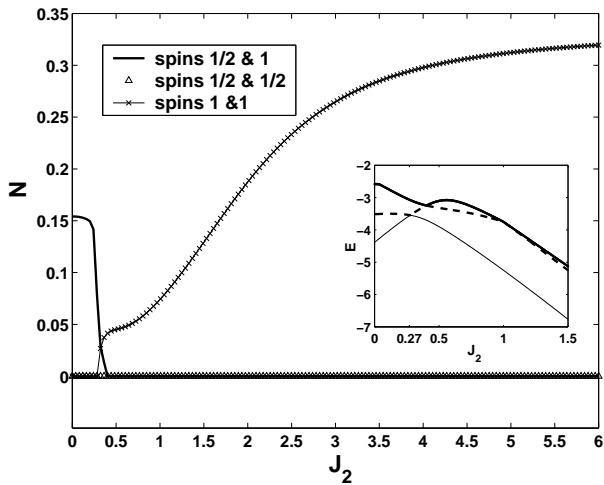}
\caption{Negativity  versus the exchange interaction $J_2$ at the
temperature $T=0.02$ in six-spin system. The ground and the first
two excited energy levels versus the $J_2$ are inserted in the
plot.}
\end{figure}
\begin{figure}
\includegraphics[width=0.45\textwidth]{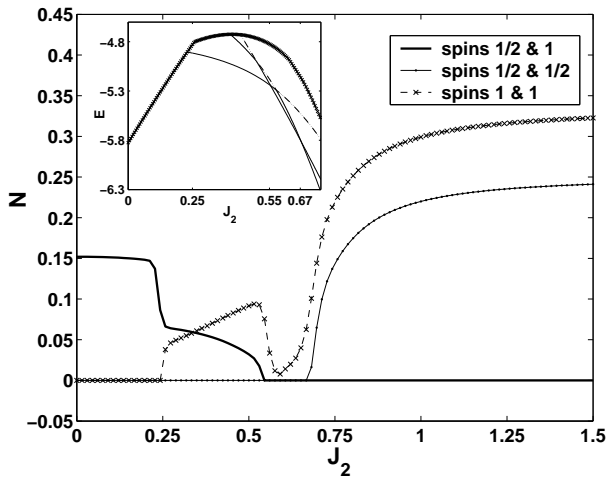}
\caption{Negativity  versus the exchange interaction $J_2$ at the
temperature $T=0.02$ in eight-spin system, The ground and the
first three excited energy levels versus the $J_2$ are inserted.}
\end{figure}

In Fig.~9,  we plot the negativity versus $J_2$ for the case of
six spins at a lower temperature. The negativity ${\cal
N}_{1/2,1}$ behaves as a decreasing function of $J_2$. It
decreases to zero at about $J_2=0.27$, which is the special point
corresponding to the energy level crossing. On the contrary,
around the point $J_2=0.27$, ${\cal N}_{1,1}$ jumps up to a
nonzero value, and then increases gradually until approaching the
limit about ${\cal N}_{1,1}=0.33$. This behavior is quite
different from that in the four-spin model.

We also see that ${\cal N}_{1/2,1/2}$ is zero all the time. It can
be understood as follows. In the six-site system there are three
spin halves with the NNN interaction. Even for a pure homogeneous
three-qubit system, there is no entanglement between two spins,
irrespective of the strength of the exchange
interactions~\cite{M_Three}. Now, in addition to the interaction
among three spin halves, there are also interaction between spin
halves and spin ones. So, it is reasonable that the entanglement
between two spin halves is zero.

The negativity versus $J_2$ in the eight-site case is shown in
Fig. 10. After the first sharp jump to a value (not zero) at about
$J_2=0.25$, ${\cal N}_{1/2,1}$ goes down to zero gradually. At
approximately $J_2=0.55$, the negativity is zero.  On the
contrary,  the negativity ${\cal N}_{1,1}$ jumps up at about
$J_2=0.25$, and then goes up gradually and almost linearly until
$J_2$ reaches about $0.55$. Then there happens a sharp decrease to
nearly zero, and after that it begins to increase gradually. The
negativity ${\cal N}_{1/2,1/2}$ keeps zero until $J_2$ reaches
about $0.67$, and then it goes up until reaches a steady value.

In Fig.~10, we find that the three kinds of negativity exhibit
different properties. From figure inserted, i.e., the energy
levels of the eight-spin system, we can see that in the region
from about $J_2=0.55$ to $J_2=0.67$ the first excited energy is
quite close to the ground energy. It is known that the energy
level crossing can greatly affect the entanglement. Here, the two
close energy levels also play an important role in the behavior of
entanglement. The approximate degenerate energy levels may
remarkably change the probability distribution even at a very low
temperature, thereby affect the negativity.
\subsubsection{Entanglement versus $J_2$ and $T$}
\begin{figure}
\includegraphics[width=0.45\textwidth]{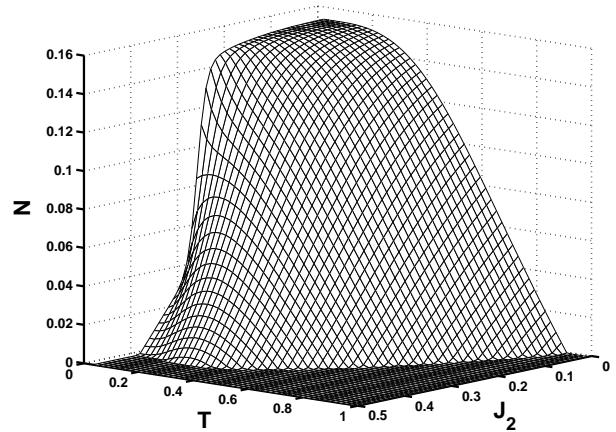}
\caption{Negativity ${\cal N}_{1/2,1}$ versus both the exchange
interaction $J_2$ and the temperature in the six-spin system.}
\end{figure}
\begin{figure}
\includegraphics[width=0.45\textwidth]{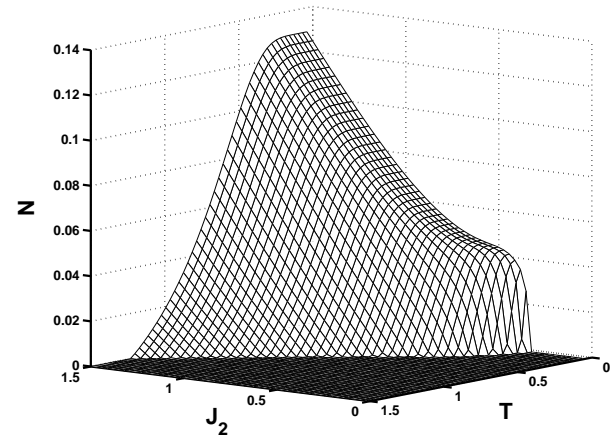}
\caption{Negativity ${\cal N}_{1,1}$ versus both the exchange
interaction $J_2$ and the temperature in six-spin system.}
\end{figure}

Now, we present the entanglement versus $J_2$ and $T$ in the
six-spin model. The NN negativity ${\cal N}_{1/2,1}$ is shown in
Fig.~11. {}On the $J_2-T$ plane, similar to the four-spin case,
the $J_\text {2th}$ does not behave as a monotonous function of
$T$ and it reaches its maximum at about $T=0.27$. From the figure,
we may find that the entanglement only exists in the region
approximately $T<0.925$ and $J_2<0.418$.  The strong NNN
interaction and thermal fluctuation will suppress the NN
entanglement to zero.

We do not plot the negativity ${\cal N}_{1/2,1/2}$ as it is zero
all along for any $J_2$ and $T$. We give the NNN negativity ${\cal
N}_{1,1}$ in Fig.~12. In the region of $J_2<0.282$, there is no
negativity ${\cal N}_{1,1}$ at any temperature. The threshold
$J_\text {2th}$ is increased by the increasing temperature,
similar to the four-spin case.

\section{conclusion}
In conclusion, we have studied the entanglement properties of the
(1/2,1) mixed-spin systems described by the Heisenberg model. In
the systems only with NN exchange interactions, for two-spin and
three-spin cases analytical results of the negativity have been
obtained, which facilitate our discussions of entanglement. The
analytical expression of threshold temperature after which the
entanglement vanishes are obtained for the two-site case. For the
case of even number of particles, it has been found that the
pairwise thermal entanglement is solely determined by the internal
energy, and thus builds an interesting relation between the
microscopic quantity, entanglement, and the macroscopic thermal
dynamical function, the internal energy in the mixed-spin systems.
For the odd number of particles such as the three-site case, we
also provide a relation between the internal energy and
negativities. We have numerically studied the effects of different
finite temperature and magnetic fields on entanglement. As a
conclusion, the thermal fluctuation suppresses the entanglement,
and entanglement may change evidently at some critical points of
magnetic field.

In the systems also with NNN interactions, by applying the angular
momentum coupling method, we obtained analytical results of all
the eigenenergies of the four-spin system, based on which the
negativity of the NN spins has been obtained. We also considered
the excited-state entanglement. At finite temperature, from the
partition function, the analytical results of negativity has been
given. We have numerically studied the effects of the NNN
interaction on the NN entanglement and NNN entanglement. It is
natural to see that the NNN interaction suppresses the NN
entanglement, and enhances the NNN entanglement. We found that the
negativity between two spin ones is sensitive to $J_2$ and
displays some interesting properties. At finite temperature, the
thermal fluctuation suppresses both the NN entanglement and the
NNN entanglement. The threshold values $J_\text {2th}$ and
$T_\text {th}$ are studied in detail. The entanglement displays
some peculiar properties, which are quite different from those of
the spin-half model. These are due to inherent mixed-spin
character of our system. It is more interesting to study
entanglements in other mixed systems and explore some universal
properties, which are under consideration.

\appendix
\section{Two-site Heisenberg model with a magnetic field}
The Hamiltonian of the two-site Heisenberg model with a magnetic
field is written explicitly as
\begin{equation}
H_1=s_{1x}\otimes S_{2x}+s_{1y}\otimes S_{2y}+s_{1z}\otimes
S_{2z}+B(s_{1z}+S_{2z}).\nonumber
\end{equation}
Following the same way as the discussions of subsection II.A, the
density matrix of the thermal state is given by Eq.~(\ref{rho})
with the matrix elements
\begin{align}
a_1=&a_6e^{3\beta B}=e^{\frac{\beta}{2}(3B-1)},\nonumber\\
a_5=&a_2e^{\beta B}= \frac{1}3e^{\frac{\beta
B}{2}}\Big(e^\beta+2e^{-\frac{\beta}2}\Big),
\nonumber\\
a_4=&a_3e^{\beta B}=\frac{1}3e^{\frac{\beta
B}{2}}\Big(2e^\beta+e^{-\frac{\beta}2}\Big),
\nonumber\\
b_2=&b_1e^{\beta B}=-\frac{\sqrt{2}}3e^{\frac{\beta
B}{2}}\Big(e^\beta-e^{-\frac{\beta}2}\Big) ,\nonumber\\
\end{align}
and the partition function
\begin{eqnarray}
Z&=&e^{\frac{\beta}{2}(3B-1)}+e^{-\frac{\beta}{2}(3B+1)}\nonumber\\
&+&e^{\frac{1}{4}\beta}\cosh\big(\frac{3\beta}{4}\big)
\cosh\big(\frac{\beta B}2\big).
\end{eqnarray}
Having obtained the analytical expressions of the matrix, we
directly obtain the negativity after substituting the matrix
elements to Eq.~(\ref{N}).

\end{document}